\documentclass[twocolumn,preprintnumbers,prd,superscriptaddress,groupedaddress,aps]{revtex4-2}
\usepackage{lmodern}
\usepackage[T1]{fontenc}
\usepackage{inputenc,psfrag,graphicx,epsfig,amsmath,amssymb,amsfonts,bbold,bm,manfnt,color,feynmp,caption,makeidx,hyperref}
\usepackage{slashed}
\usepackage{nicefrac}
\usepackage{braket}
\usepackage{cleveref}
\usepackage{subcaption}
\usepackage{tikz}
\usetikzlibrary{decorations.pathmorphing,patterns,quotes,angles}
\usetikzlibrary{intersections,through,positioning}

\def\beq{\begin{equation}}
\def\eeq{\end{equation}}
\def\beqa{\begin{eqnarray}}
\def\eeqa{\end{eqnarray}}

\newcommand{\bea}{\begin{eqnarray}}
\newcommand{\eea}{\end{eqnarray}}
\newcommand{\bean}{\begin{eqnarray*}}
\newcommand{\eean}{\end{eqnarray*}}

\newcommand{\Disc}[0]{\operatorname{Disc}}

\def\eps{\epsilon}

\def\half{\frac{1}{2}}

\newcommand{\rd}{\mathrm{d}}

\newcommand{\cF}{\mathcal{F}}
\newcommand{\cU}{\mathcal{U}}

\newcommand{\cI}{\mathcal{I}}

\newcommand{\cC}{\mathcal{C}}

\begin{document}

\title{\boldmath
Generalized Cuts of Feynman Integrals in Parameter Space
\unboldmath}

\author{Ruth Britto}
\affiliation{Institute for Advanced Study, Einstein Drive, Princeton NJ 08540, USA}
\affiliation{School of Mathematics and Hamilton Mathematics Institute, Trinity College, Dublin 2, Ireland}

\begin{abstract}

\noindent

We propose a construction of generalized cuts of Feynman integrals as an operation on the domain of the Feynman parametric integral. 
A set of on-shell conditions removes the corresponding boundary components of the integration domain, in favor of including a boundary component from the second Symanzik polynomial. Hence integration domains are full-dimensional spaces with finite volumes, rather than being localized around poles. 
As initial applications, we give new formulations of maximal cuts, and we provide a simple derivation of a certain linear relation among cuts from the inclusion-exclusion principle.


\end{abstract}

\maketitle


Generalized cuts of Feynman integrals are discontinuities around their Landau singularities. The operation of cutting a Feynman integral, or by extension a physical scattering amplitude, has been widely used for computation in on-shell or unitarity methods, whether from a reconstruction based directly on discontinuities, as a seed for finding solutions of their differential equations, or through pattern-matching based on their functional structure.

Cuts are typically defined by constraining a subset of the internal momenta to be on-shell, and their computation can be carried out by taking residues at that locus. Here, we outline a complementary approach in terms of Feynman parameters. The Landau conditions for the presence of a singularity are formulated at first in terms of both loop momenta and Feynman parameters. Focusing on the loop momenta gives the customary on-shell conditions. Focusing instead on the Feynman parameters, we interpret the conditions in terms of changing the boundaries of the domains of integration.
%
Specifically, the coordinate hyperplanes of the parameters corresponding to cut edges are eliminated as boundaries, and the second Symanzik polynomial $\cF$ will be used in their place.
We justify this interpretation in terms of the Landau conditions, and with reference to a physical notion of discontinuities. Maximal cuts are obtained by integrating over a region bounded only by $\cF=0$. We find a new expression for maximal cuts at one loop and illustrate the possibility of finding multiple solutions beyond one loop.

In this framework, we can derive a linear relation among cut and uncut one-loop integrals in a very simple way, by an inclusion-exclusion argument on regions with the appropriate boundaries. We conclude with prospects for extending this argument to multiloop integrals, along with various suggestions for further exploration.

\section*{Feynman-parametric integrals and their cuts}

\paragraph*{Review of Feynman parameters.}
An $L$-loop Feynman integral with $E$ internal edges in $D$ spacetime dimensions is given in its momentum representation as \footnote{We suppress the factor $e^{L \gamma_E \eps}$ that is conventional in dimensional regularization.} 
\beq\label{eq:fint}
\cI = 
\int \left(\prod_{\ell=1}^L \frac{\rd^D k_\ell}{i\pi^{D/2}}\right)
\frac{B}{\prod_{i=1}^E A_i^{\nu_i}}\,.
\eeq
The numerator $B$ is an arbitrary polynomial in the kinematic variables. 
Here we treat singularities in generality, so the numerator plays no role, and we omit it henceforth.
The inverse propagators in the denominator
take the form
\beq
\label{eq:Ai}
A_i= m_i^2-q_i^2-i0\,,
\eeq 
with masses $m_i$ and momenta $q_i$ that are linear combinations of the external momenta and loop momenta $k_\ell$,
and they are raised to integer powers $\nu_i$.

Feynman parameters are the integration variables introduced by rewriting the denominators in the following form:
\begin{multline}
\label{eq:ftrick}
\frac{1}{\prod_{i=1}^E A_i^{\nu_i}}
= \frac{\Gamma(\nu)}{\prod_{i=1}^E \Gamma(\nu_i)}
\int_{\alpha_i \geq 0} \rd^E\alpha\, \\
\frac{\delta\left(1-\sum_{i \in S} \alpha_i\right)\,
\left(\prod_{i=1}^E \alpha_i^{\nu_i-1} \right)}
{\left( \sum_{i=1}^E \alpha_i A_i \right)^{\nu} }\,,
\end{multline}
where $\nu =\sum_{i=1}^E \nu_i $, and $S$ is any nonempty subset of $\{1,\ldots,E\}$.

\paragraph*{Review of Landau equations and cuts.}
The Landau equations \cite{Landau:1959fi,SMatrix} are a set of necessary conditions 
for a singularity to occur in the integral.
For each $i \in \{1,\ldots,E\}$,
\beq
\label{eq:landau-1}
\alpha_i A_i = 0\,,
\eeq
and for each closed loop of the graph, $\ell=1,\ldots,L$,
\beq
\label{eq:landau-2}
\sum_{i=1}^E \alpha_i \frac{\partial A_i}{\partial k_{\ell}}=0\,.
\eeq

Notice that the conditions of \cref{eq:landau-1} factorize, so for each Landau singularity, there is a subset $J \subseteq \{1,\ldots,E\}$ such that
\beq\label{eq:dual}
A_j=0 \textrm{~for~} j \in J\,, \qquad  \alpha_k=0 \textrm{~for~} k \notin J\,. 
\eeq
 The conditions $A_j=0$ are {\em on-shell} conditions constraining the loop momenta. We say that the propagators indexed by $j \in J$ are {\em cut}.
In the momentum-space integral \cref{eq:fint}, these conditions can be imposed by replacing the $A_j$ factors with corresponding delta functions \cite{Cutkosky:1960sp}, or by taking residues \cite{teplitz}. 
The result of this operation on the Feynman integral is a discontinuity called a {\em cut integral}, 
denoted by $\cC_J I$.

In Feynman parameter space, we  focus on the second equation of \cref{eq:dual}, the complement of the usual on-shell conditions, as the defining condition for the cut $\cC_J I$.
After rewriting the integral of \cref{eq:fint} with the identity \cref{eq:ftrick}, one can integrate the loop momenta, leaving an  integral  over the Feynman parameters $\alpha_i$ as follows:
\begin{multline}
\label{eq:FP-gen}
\cI = \frac{\Gamma\left(\nu-\frac{LD}{2}\right)}{\prod_{i=1}^E \Gamma(\nu_i)}
\int_{\alpha_i \geq 0} \rd^E\alpha\, \\
\delta\left(1-\sum_{i \in S} \alpha_i\right)\,
\left(\prod_{i=1}^E \alpha_i^{\nu_i-1} \right)
\,\frac{\cU^{\nu-(L+1)D/2}}
{\cF^{\nu-LD/2}}\,,
\end{multline}
where $\cU$ and $\cF$ are 
the first and second Symanzik polynomials. 

The Landau equations can  be recast in terms of Feynman parameters as 
\beq
\label{eq:landauFP}
\alpha_k=0 \textrm{~for~} k \notin J\,, \qquad \frac{\partial \cF}{\partial \alpha_j}=0  \textrm{~for~} j \in J\,,
\eeq
provided that $\cU \neq 0$ \footnote{For subleading singularities with $J \neq E$ and $\cU=0$, the second equation is satisfied trivially but can be refined to expose the singularity. See for example \cite{SMatrix,NakanishiGraphTheoryFeynman,Boyling1968AHA,Klausen:2021yrt,Hannesdottir:2022bmo}}.

\paragraph*{Cuts and contours.}
The conditions $\alpha_k = 0$ appear in the boundary of the domain of integration of \cref{eq:FP-gen}. It was noted in \cite{Abreu:2017ptx} that the Landau conditions treat singular surfaces and boundaries on the same footing. 

%

Our proposal is to interpret generalized cut integrals through modification of the integration contours of \cref{eq:FP-gen}, eliminating the hyperplanes $\{\alpha_k=0\}_{k \notin J}$ as boundaries, in favor of including the singular surface $\cF=0$ as a boundary.

This interpretation is supported by the second Landau equation.  At a singularity where $\cU \neq 0$, the second equation of \cref{eq:landauFP} indicates that the gradient of $\cF$ is normal to the coordinate hyperplanes associated to the cut. In other words, the surface $\cF=0$ is tangent to this set of hyperplanes. So it is plausible that crossing the singularity creates a new region bounded by $\cF=0$ and the hyperplanes $\{\alpha_k=0\}_{k \notin J}$, which can be taken as a domain of integration for defining $\cC_J \cI$ as a parametric integral. We will make this idea more precise in the cases studied in the following sections.

The idea of using full-dimensional integration contours 
is related to constructing a 
basis of the homology group associated to a
generalized hypergeometric function \cite{AomotoKita}. 
The integration contours are most properly constructed as twisted cycles, meaning that they include information about the choice of the Riemann sheet with respect to the multivalued integrand. 
In practice, to obtain the functions that interest us, it is sufficient to integrate in the usual way, employing dimensional regularization as needed.

\paragraph*{Notation and conventions.}
We use the delta function in the parametric integral to restrict our attention to the hyperplane $H_S$ defined by $1-\sum_{i \in S} \alpha_i=0$.
%
We denote the coordinate hyperplanes by 
$H_i \equiv \{\alpha_i=0\}$, 
the restriction of $\cF=0$ to $H_S$ by $F$, and 
%
the integration region of the uncut integral by $\Delta.$ 
%
Where integral formulas are given, it is to be understood that parameters such as $D$ are chosen to ensure convergence, for example through 
dimensional regularization.
%

\section*{Discontinuities and contours}

Let us examine the discontinuity of the Feynman integral in a general kinematic variable denoted by $s$. Here we interpret the discontinuity 
as the difference in the value of a function in approaching the real axis from opposite sides. More formally, a discontinuity is the change resulting from analytic continuation around a Landau variety, which can be treated with Picard-Lefschetz theory \cite{fotiadi,teplitz,PhamBook,Abreu:2017ptx,Hannesdottir:2022xki,Berghoff:2022mqu}. 

To see explicit discontinuities in mass and momentum invariants, we refer to the graphical definitions of the Symanzik polynomials, 
\begin{align}\begin{split}
\cU
 & = 
 \sum_{T\in {\mathcal T}_1} 
     \prod_{e_i\notin T} \alpha_i\,,
 \\
\cF
 & = {\mathcal U} \sum_{i=1}^{E} \alpha_i m_i^2 
 - \sum\limits_{(T_1,T_2)\in {\mathcal T}_2} 
     \left( \prod_{e_i\notin (T_1,T_2)} \alpha_i \right) 
q_{(T_1,T_2)}^2 \,,
\end{split}\end{align}
where 
${\mathcal T}_1$ is the set of spanning trees of the Feynman graph  and ${\mathcal T}_2$ is the set of spanning 2-forests with connected components $T_1$ and $T_2$, in which $q_{(T_1,T_2)}$ is the momentum flowing bewteen the connected components.
Since the kinematic dependence is only in the polynomial $\cF$, the discontinuity operation acts only on that factor. 

For simplicity, suppose that there exists a 
region in the space of kinematic variables in which  $\cF$ is positive throughout the integration region $\Delta$,
called a Euclidean region.
Thus $\Delta$ is contained in the interior of $\cF>0$. 

The $i0$ prescription in \cref{eq:Ai} can be associated to the kinematic invariants, and thus to $\cF-i0$,  so that the discontinuity of 
$\cF^{\lambda} $ is
\begin{align}\begin{split}
\label{eq:disc-gen}
 \Disc_s\left[\cF^{\lambda} \right]
&= \left(\cF -i0\right)^{\lambda} - \left(\cF +i0\right)^{\lambda} \\
&=  -\theta[-\cF] \left[-\cF\right]^{\lambda} 
 2 i \sin(\pi \lambda)\,, 
\end{split}\end{align}
where $\theta$ is the unit step function. 
Inserting this expression into the integral, we see concretely from the factor $\theta[-\cF]$ that $\cF=0$ has effectively become a boundary of the integration region. 

\begin{figure}
	\begin{subfigure}{.2\textwidth}
		\centering
		\includegraphics[height=3cm]{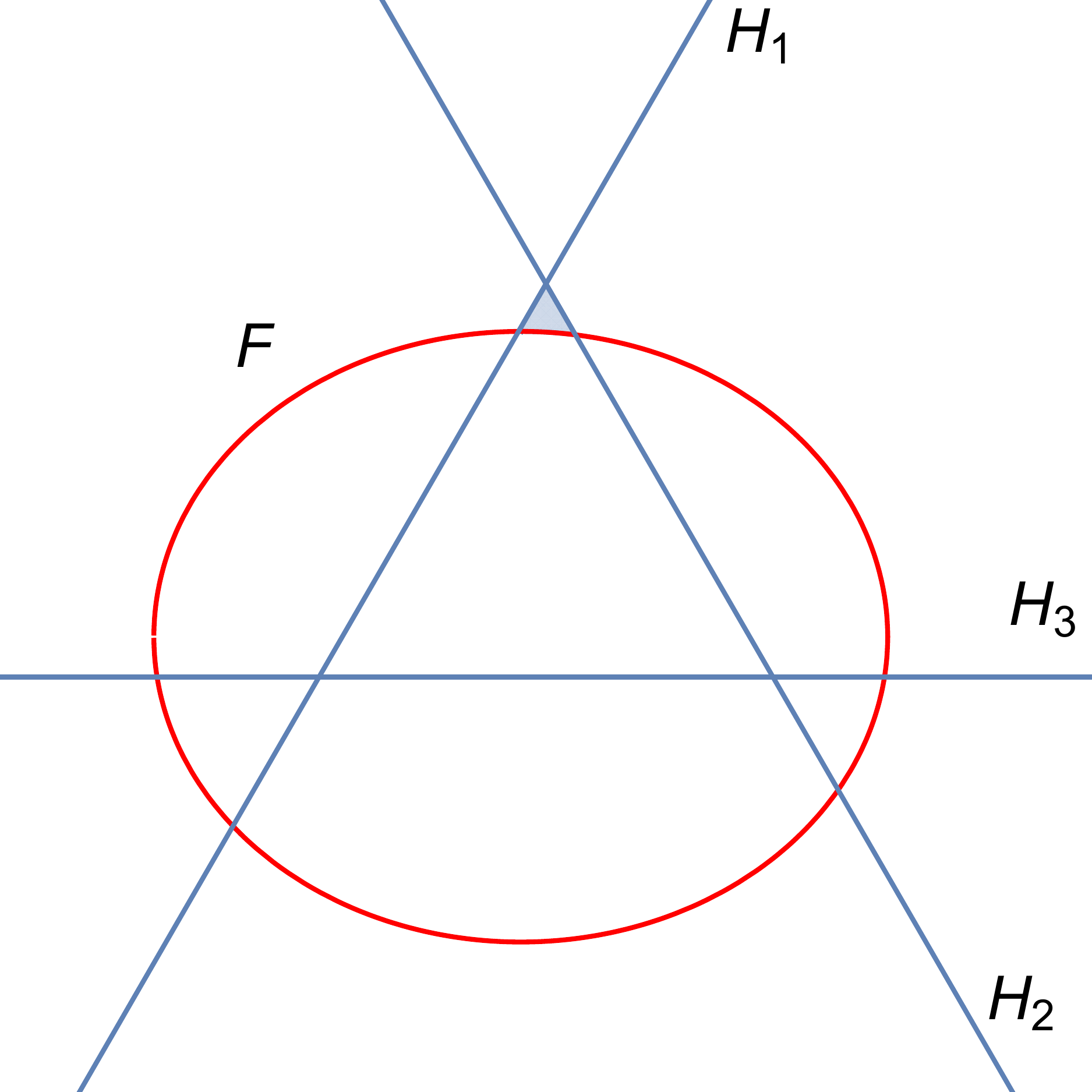}  
		\caption{Discontinuity in $m_3^2$.}
		\label{fig:mass-disc}
	\end{subfigure}
\begin{subfigure}{.2\textwidth}
	\includegraphics[height=3cm]{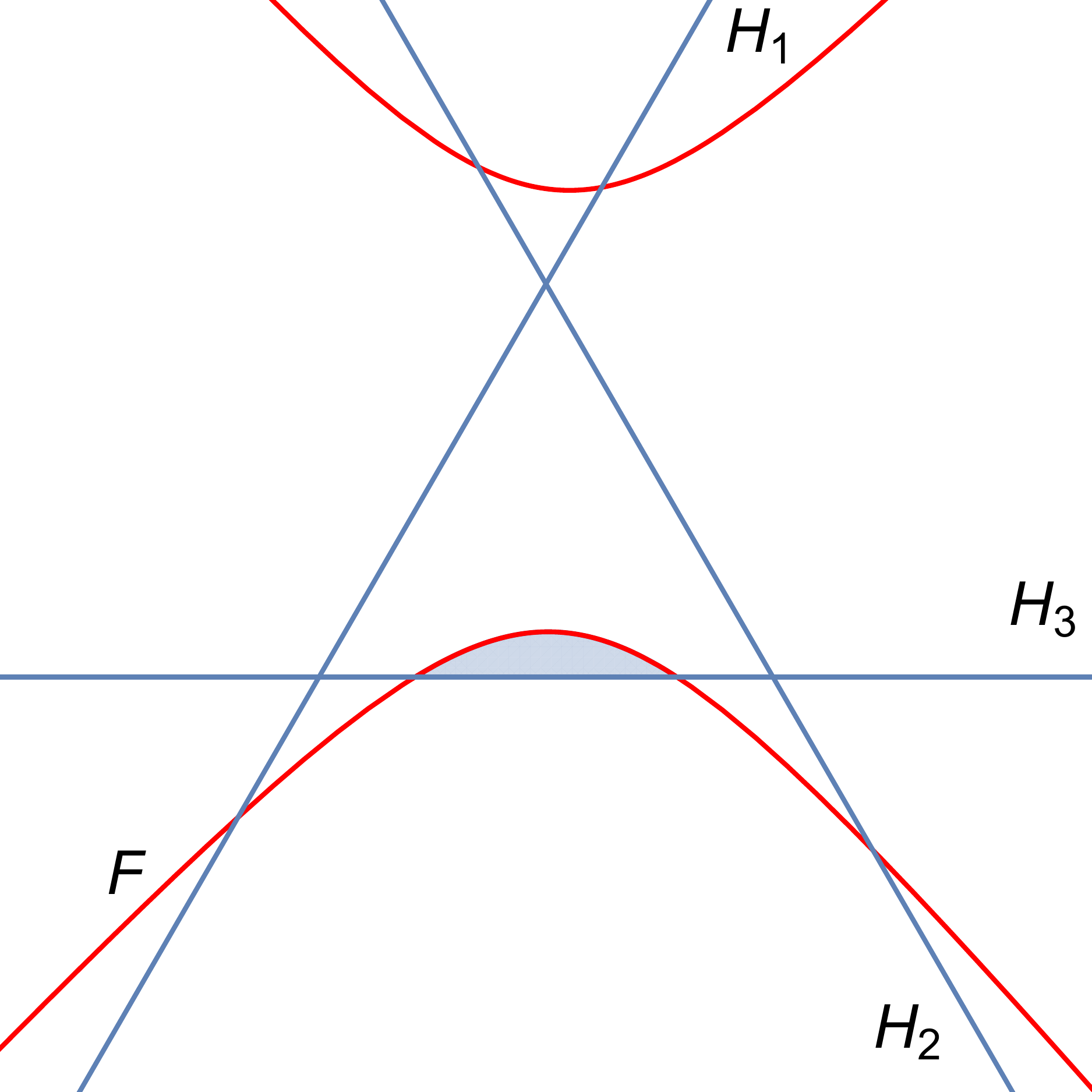}
	\caption{Discontinuity in $p_3^2$.}
		\label{fig:mom-disc}
\end{subfigure}
	\caption{Integration domains of mass and momentum discontinuities of the scalar triangle.}
	\label{fig:discs}
\end{figure}

\paragraph*{Mass discontinuities.}
The effect of a discontinuity operation on the integration region is most clearly seen in varying the mass of a single edge.
Consider the example of a scalar triangle.
See \cref{fig:mass-disc}. 
As  the value of  $m_3^2$ is analytically continued from positive to negative values, $F$ crosses into the integration region at the point $H_{1}\cap H_2$, eliminating $H_3$ as a boundary. 

\paragraph*{Momentum discontinuities.} 
Discontinuities in momentum invariants are less elementary. In multiloop integrals, it may be necessary to combine multiple cuts (multiple choices of cut edges) to reproduce the full discontinuity. Each cut diagram has a separate integration region. 
In our example of the scalar triangle,  
as $p_3^2$ crosses the normal threshold,
where $F$ is tangent to $H_3$, the integration region bounded only by $F$ and $H_3$ appears, seen in \cref{fig:mom-disc}.
\footnote{At the anomalous threshold, $F$ is tangent to $H_3$ away from $\Delta$.}.
%

\paragraph*{Relation of discontinuities to cuts.}
As noted above, discontinuities exist in a more general sense captured by Picard-Lefschetz theory, and they are represented by the full set of cuts even when the corresponding Landau singularities may not have solutions in real kinematics. For example, the functions expressing  mass and momentum discontinuities 
may be analytically continued to other kinematic regions, or equivalently, computed directly in other regions via cut integrals. 
Cuts can be defined modulo $i\pi$ in order to preserve invariance under analytic continuation.
In the present discussion, the analytic continuation of cuts is derived from the analytic continuation of $\cF$. 
For cuts to agree with discontinuities, we should normalize the contours by the factor $-2i\sin(\pi(LD/2-\nu))$ from \cref{eq:disc-gen}, where we have taken the value of $\lambda$ from \cref{eq:FP-gen}. If $D$ is an even integer minus $2\epsilon$, this factor is of order $\epsilon$.

\begin{figure}
	\begin{subfigure}{0.23\textwidth}
		\centering
		\includegraphics[width=0.8\linewidth]{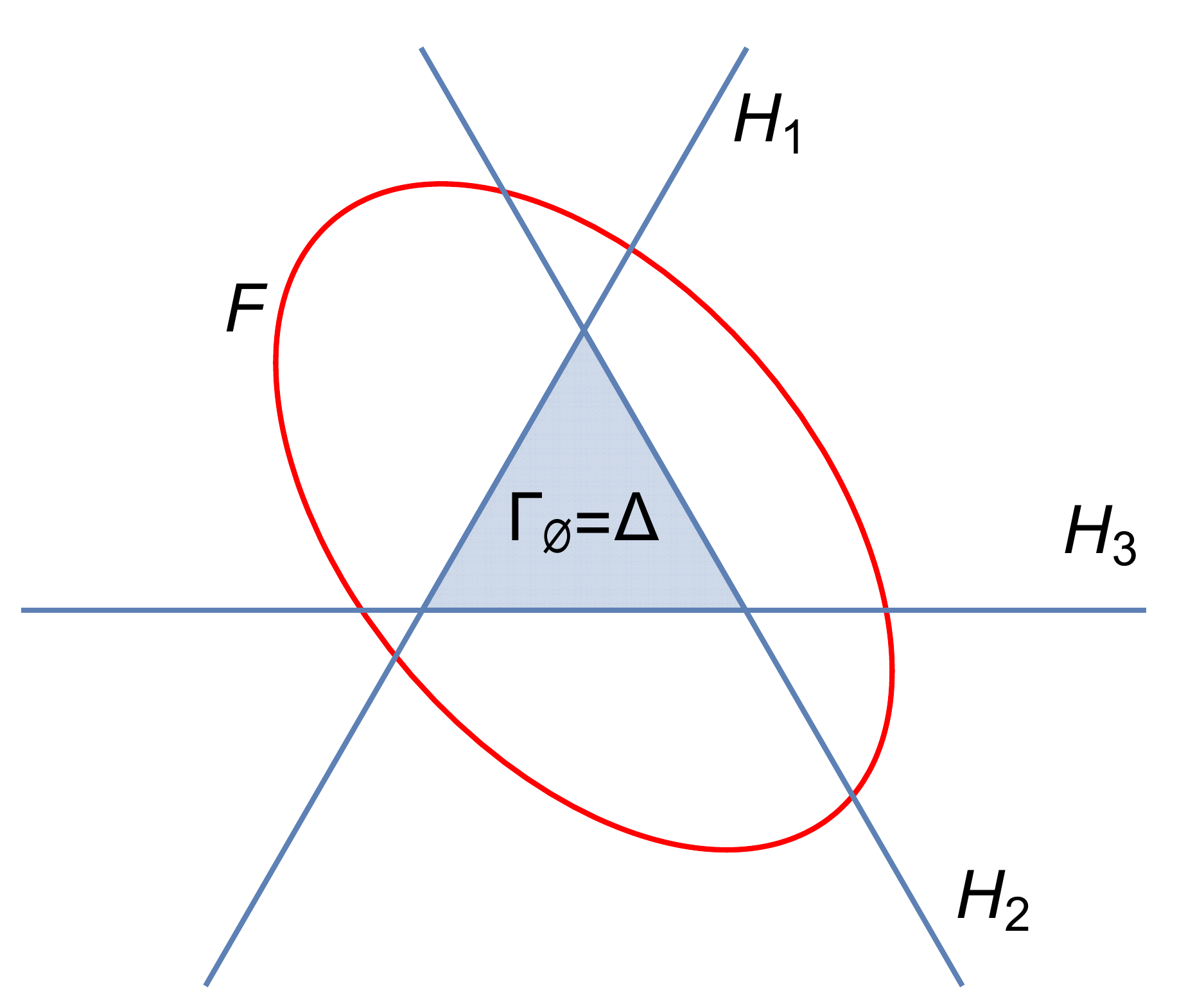}  
		\caption{$\Gamma_\emptyset$}
	\end{subfigure}
	\begin{subfigure}{0.23\textwidth}
		\centering
		\includegraphics[width=0.8\linewidth]{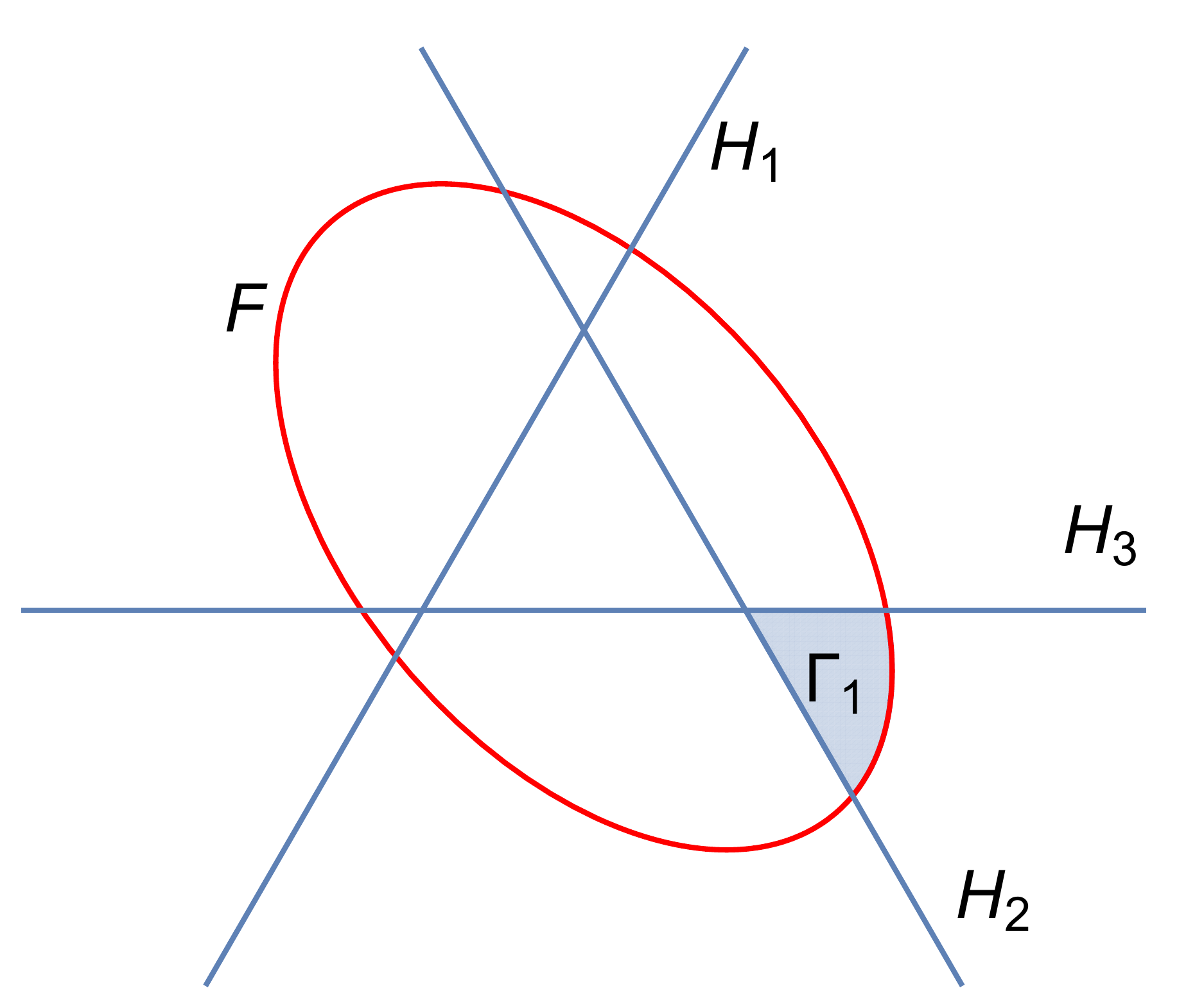} 
		\caption{$\Gamma_1$} 
	\end{subfigure}
\newline
	\begin{subfigure}{.23\textwidth}
		\centering
		\includegraphics[width=0.8\linewidth]{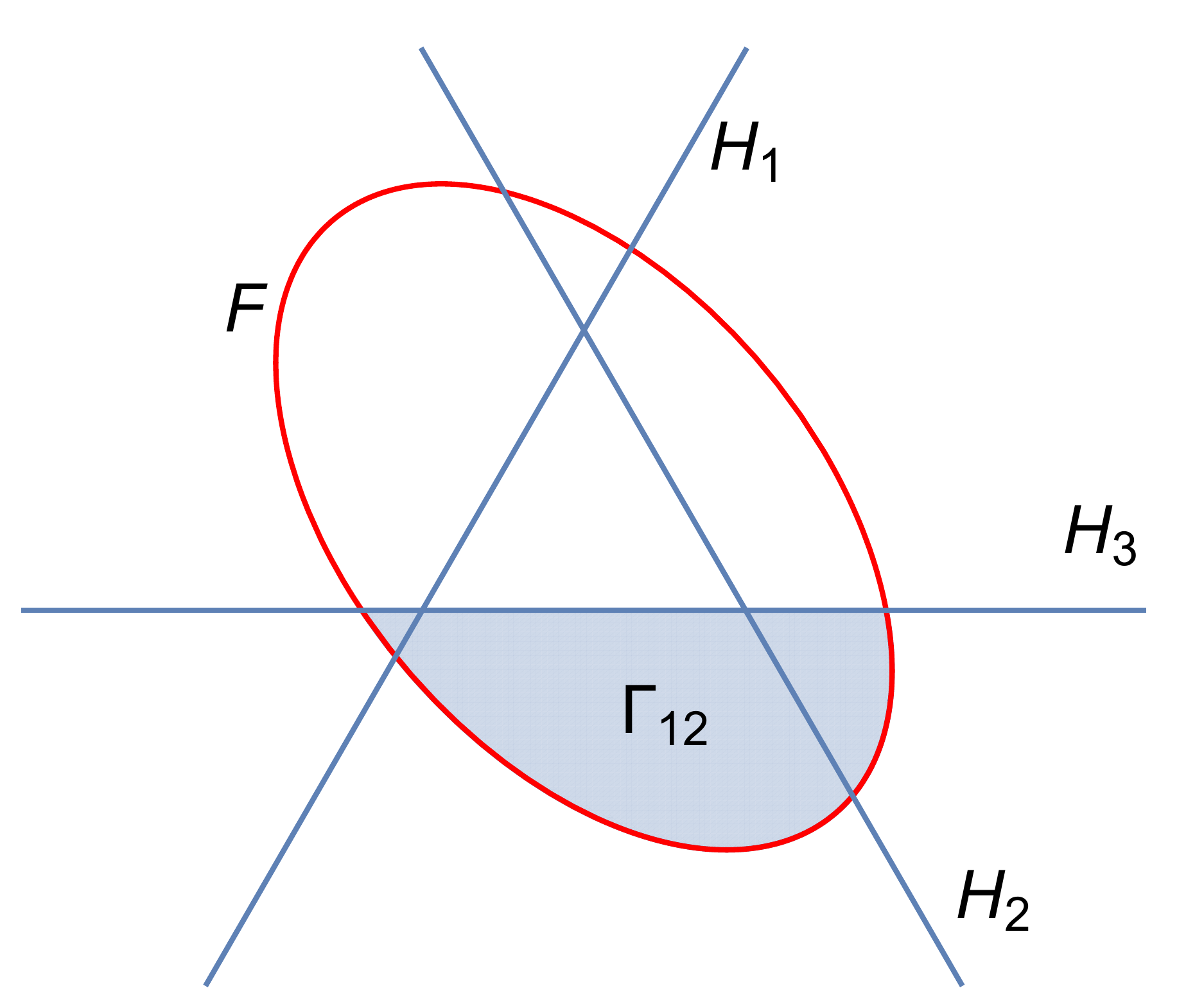}  
		\caption{$\Gamma_{12}$}
	\end{subfigure}
\begin{subfigure}{.23\textwidth}
	\includegraphics[width=0.8\linewidth]{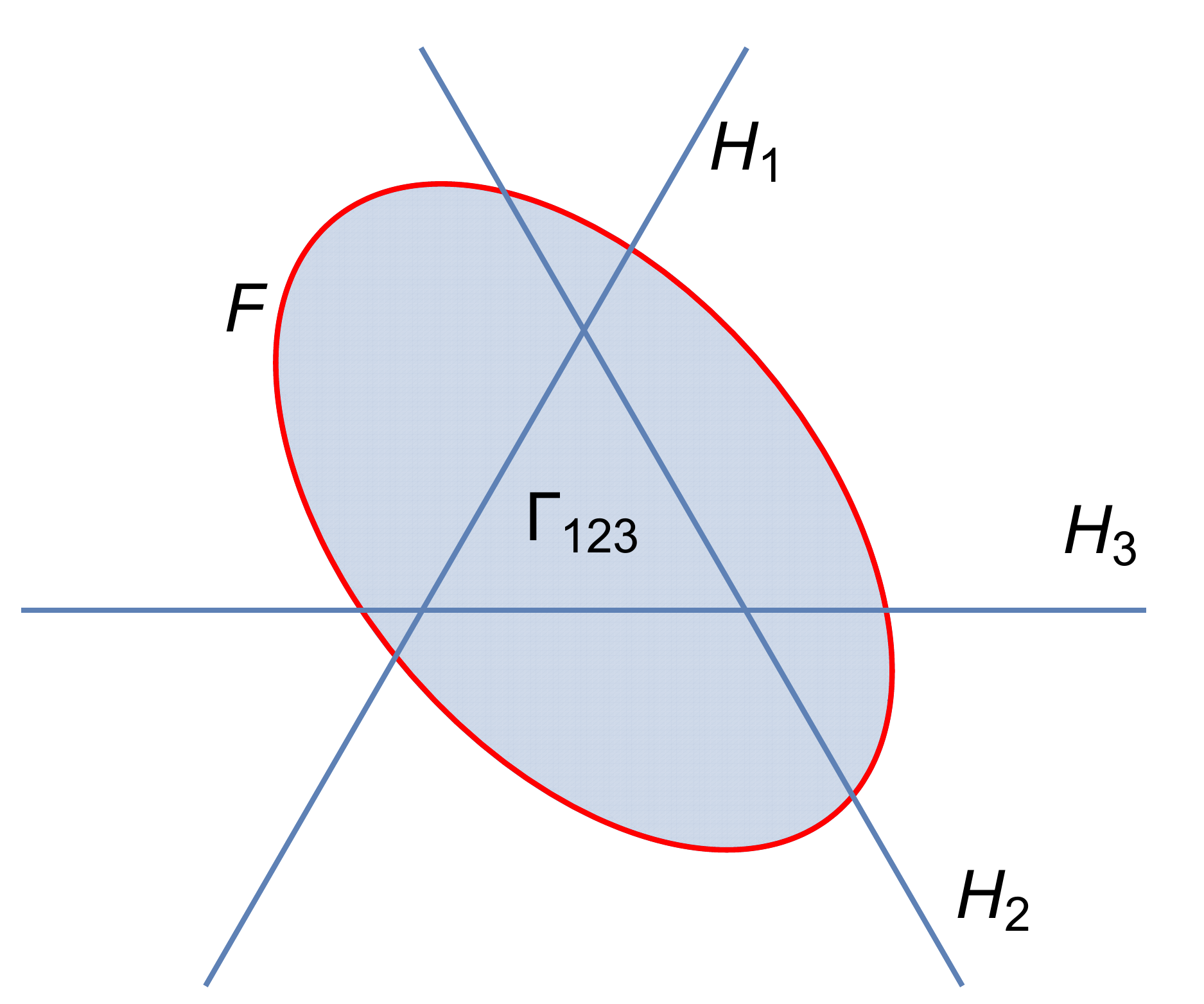}
	\caption{$\Gamma_{123}$}
\end{subfigure}
	\caption{Integration domains of an $n$-point one-loop integral in Euclidean kinematics, shown for $n=3$. 
	}
	\label{fig:tri-regions}
\end{figure}

\paragraph*{One-loop cuts.}
For example, in a one-loop integral, integration domains can be continued back to the Euclidean region, as shown in \cref{fig:tri-regions}.
The cut integral $\cC_J \cI$ can be interpreted as the Feynman integral over the integration domain $\Gamma_{J}$ bounded by $F$ and $\{H_k\}_{k \notin J}$. 
In particular, the domain associated to the maximal cut is $\Gamma_{[n]}$, which is bounded only by  $\cF =0$.
The choice of boundaries does not define the region uniquely. The tangency of $F$ to the hyperplanes seen from the second Landau equation implies that with the exception of $\Gamma_{[n]}$, we should take the integration domain of a cut integral to exclude $\Delta$ \footnote{This choice ensures that cuts agree with physical discontinuities. A different choice would give a linear combination of cuts with additional on-shell conditions. For the maximal cut, the validity of our choice can be seen by consistency with the case of $n=2$, where the maximal cut is also a momentum discontinuity.}. 
%

If cuts are defined  such that they agree with discontinuities up to sign, and we assume single powers of propagators, then  if $J \neq \emptyset$  and $D$ is an integer minus $2\epsilon$, 
\begin{align}\begin{split}\label{eq:1loop-cuts-contours}
\cC_J \cI = & \frac{\sin(\pi\eps)}{\pi} \,\Gamma\left(n-\frac{D}{2}\right)\,
\int_{\Gamma_{J}}\,
\omega\,
\cF^{\frac{D}{2}-n}\,,
\end{split}\end{align}
where $\omega$ is the volume form induced on $H_S$ \footnote{This formula differs from the conventions of \cite{Abreu:2017ptx} in that we have not removed overall factors of $2\pi i$.}. 

\section*{Maximal Cuts}

We  carry out some more checks of our proposal by evaluating maximal cuts in a few simple cases. A maximal cut is defined by 
setting 
$A_j=0$ for all $j$. None of the coordinate hyperplanes remain as boundaries of the integration region; the region for the maximal cut contour is bounded only by $F$.
Maximal cuts of multiloop integrals are not necessarily unique. 

\paragraph*{One-loop integrals.}
The maximal cut of a scalar one-loop $n$-point integral with unit powers of propagators can be computed in generality. 
In \cref{eq:FP-gen}, set $\nu_i=L=1$, and take $S=[n].$
The graph polynomials are 
$
\cU = \sum_{j=1}^{n}\alpha_j $, 
which is set to $1$ by the delta function,
and 
$\cF = \sum_{i,j=1}^{n} Y_{ij} \alpha_i \alpha_j \,,
$
where the $Y_{ij}$ are entries of the modified Cayley matrix,
$
Y_{ij} = \half\left[m_i^2 + m_j^2 - (q_i - q_j)^2\right].
$

It is convenient to introduce the  $n$-vectors $\alpha=(\alpha_1,\ldots,\alpha_n)^T$ and $e=(1,\ldots,1)^T$. Let $Y$ be the symmetric matrix with entries $Y_{ij}$. Then  
$\cU = e^T \alpha$ and $\cF = \alpha^T Y \alpha$, and the maximal cut integral is 
\begin{multline}
\int_{\alpha^T Y \alpha \geq 0}\,
\rd^n\alpha\, \delta(1-e^T \alpha)
\left(\alpha^T Y \alpha\right)^{\frac{D}{2}-n}
= \\
\frac{1}{\sqrt{\det Y}} 
\left(e^T Y^{-1} e\right)^{\frac{n-D}{2}}
\pi^\frac{n-1}{2}
\frac{\Gamma\left(\frac{D}{2}-n+1\right)}{\Gamma\left(\frac{D-n+1}{2}\right)}\,,
\end{multline}
which agrees with the analytic result of \cite{Abreu:2017ptx}.

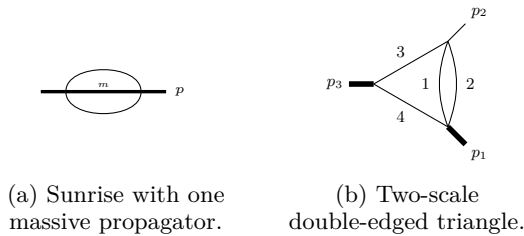
\begin{figure}[h]
	\begin{subfigure}{0.22\textwidth}
		\centering
	\raisebox{0.75\height}{
		\begin{tikzpicture}
		\coordinate (G1) at (0,0);
		\coordinate (G2) at (1,0);
		\coordinate (H1) at (-1/3,0);
		\coordinate (H2) at (4/3,0);
		\coordinate (I1) at (1/2,1/2);
		\coordinate (I2) at (1/2,1/8);
		\coordinate (I3) at (1/2,-1/2);
		\coordinate (I4) at (1/2,-1/8);
		\coordinate (J1) at (1,1/4);
		\coordinate (J2) at (1,-1/4);
		\coordinate (K1) at (1/2,-1/8);
		\draw (G1) [line width=0.45 mm] -- (G2);
		\draw (G1) to[out=90,in=90] (G2);
		\draw (G1) to[out=-90,in=-90] (G2);
		\draw (G1) [line width=0.45 mm] -- (H1);
		\draw (G2) [line width=0.45 mm]-- (H2);
		\node at (H2) [right=0 mm of H2] {\tiny$p$};
		\node at (K1) [above=1.0 mm of K1,scale=0.7] {\tiny$m$};
		\end{tikzpicture}
		%
	}
		\caption{Sunrise with one massive propagator.}
		\label{fig:sunrise}
	\end{subfigure}
	\begin{subfigure}{0.2\textwidth}
	\centering
	\resizebox{2.5cm}{!}{
	\begin{tikzpicture}
	\coordinate (G1) at (0,0);
	\coordinate (G2) at (1,0.57735026919);
	\coordinate (G3) at (1,-0.57735026919);
	\coordinate (H1) at (-1/3,0);
	\coordinate [above right = 1/3 of G2](H2);
	\coordinate [below right = 1/3 of G3](H3);
	\coordinate (I1) at (1/2,0.28867513459);
	\coordinate [above left=1/4 and 1/8 of I1](I11);
	\coordinate [below right = 1/4 and 1/8 of I1](I12);
	\coordinate (I2) at (1/2,-0.28867513459);
	\coordinate [below left = 1/4 and 1/8 of I2](I21);
	\coordinate [above right = 1/4 and 1/8 of I2](I22);
	\coordinate (I3) at (1,0);
	\coordinate (I31) at (3/4,0);
	\coordinate (I32) at (5/4,0);
	\coordinate [above left=1/4 of G2] (J1);
	\draw (G1) -- (G2);
	\draw (G1) -- (G3);
	\draw (G2) to[out=-110,in=110] (G3);
	\draw (G2) to[out=-70,in=70] (G3);
	\draw (G1) [line width=0.75 mm] -- (H1);
	\draw (G2)  -- (H2);
	\draw (G3) [line width=0.75 mm] -- (H3);
	\node at (H1) [left=0,scale=0.7] {\small$p_3$};
	\node at (H2) [above right=0,scale=0.7] {\small$p_2$};
	\node at (H3) [below right=0,scale=0.7] {\small$p_1$};
	\node at (0.5,0.57735026919/2) [above left=0,scale=0.7] {\small$3$};
	\node at (0.5,-0.57735026919/2) [below left=0,scale=0.7] {\small$4$};
	\node at (0.7,0) [scale=0.7] {\small$1$};
	\node at (1.3,0) [scale=0.7] {\small$2$};
	\end{tikzpicture}
	%
		%
	}
		\caption{Two-scale double-edged triangle.} 
		\label{fig:dtri}
	\end{subfigure}
\newline
	\caption{Examples of two-loop graphs.}
	\label{fig:2loops}
\end{figure}

\paragraph*{Sunrise with one massive propagator.} 
For the 
scalar integral depicted in \cref{fig:sunrise},
we choose $S=\{1,2\}$ and set $\alpha_1=x,~ \alpha_2=1-x,~ \alpha_3=y$, where $\alpha_3$ is the parameter of the massive edge. 

Then the graph polynomials are $\cU = x(1-x)+y,~ \cF = y \left[m^2 y + (m^2-p^2) x(1-x)\right]$.
Integrating $\cU^{3-D/2} \cF^{D-3}$  between the two roots of $\cF(y)$ produces a  hypergeometric function, ${}_2F_1$, whose argument is $1-p^2/m^2$. The remaining $x$ dependence is simply a factor $[x(1-x)]^{(D-4)/2}$; integrating between the intersection points of the factors of $\cF=0$ at $x=0$ and $x=1$ reproduces the known discontinuity in $p^2$, as given for example in \cite{Abreu:2021vhb}.

In this example, there are two independent maximal cuts. It is interesting to note that both can be obtained by a suitable choice of contour bounded only by $F$. For example, we can obtain a second independent solution by integrating between $y=0$ and $y=\infty$, and then integrating over all values of $x$.

\paragraph*{Double-edged triangle with massless propagators and two legs off shell.}
For the scalar integral depicted in \cref{fig:dtri},
we choose $S=\{1,2\}$ and set $\alpha_1=x,~  \alpha_2=1-x,~  \alpha_3=y,~  \alpha_4=z$.  Here there is only one maximal cut.
The two factors of the polynomial $\cF=-z[p_3^2 y + p_1^2 x(1-x)]$ create four unbounded regions of integration. 
Integrating over any of them gives the same result up to normalization, which is proportional to 
 $[p_3^2(p_3^2-p_1^2)]^{(D-4)/2}$ and agrees with the known result as given for example in \cite{Abreu:2021vhb}.

\paragraph*{Remark.} Nonmaximal cuts of the integrals in \cref{fig:2loops} also agree with known results.

\section*{The decomposition theorem from the inclusion-exclusion principle}

A primary motivation of this work is the study of linear relations among cut integrals. For one-loop integrals, several such relations were analyzed in \cite{Abreu:2017ptx}. Of particular interest is the equivalence of an uncut $n$-point scalar integral to the sum of all its 1-line and 2-line cuts,
\beq \label{eq:pci}
\sum_{j=1}^n\cC_{\{j\}} \cI + \sum_{\{j,k\}\subseteq [n]}\cC_{\{j,k\}} \cI \equiv -\epsilon\, \cI \quad \textrm{mod~}i\pi\,.
\eeq
Here we are working in dimensional regularization and assuming that   $D$ is an even integer minus $2\epsilon$. 
 This relation was derived from a decomposition theorem in the homology of Feynman integrals \cite{fotiadi} showing that the singularity at infinite momentum can be exchanged for the singularities associated to cuts of edges. The singularity at infinity produces the term on the right-hand side of \cref{eq:pci}. In fact, the decomposition theorem involves the homology classes of all combinations of cut edges, not just single edges and pairs. The reason that they do not appear 
 in \cref{eq:pci} is that the relation is given modulo $i \pi$, consistent with defining cuts to be invariant under analytic continuation to different kinematics. Computed as residues or discontinutes, cuts of larger numbers of edges are accompanied by higher powers of $i\pi$ relative to the first two \footnote{These additional powers of $i\pi$ are conventionally removed through normalization when studying the functional form of cuts.}. We can now give a version of this relation that is exact in the Euclidean region, from which it can be analytically continued to other regions as desired.

The derivation follows very simply from the inclusion-exclusion principle applied to bounded regions. We refer again to \cref{fig:tri-regions}.
The set of integration domains $\Gamma_J$  with $J \neq \emptyset$, associated to cuts, is linearly independent. The domain $\Gamma_{\emptyset}=\Delta
$ associated to the uncut integral can be expressed in terms of the others by an alternating sum,
giving
\beq
\sum_{\emptyset \subset K \subseteq [n]} (-1)^{n-|K|} \Gamma_K 
=  \Gamma_{\emptyset}\,.
\eeq
Now we use this expression as an integration contour for the Feynman integral and recall \cref{eq:1loop-cuts-contours}. The sum on the left becomes the sum of all cuts times $\pi \csc(\pi \epsilon)$. 
This is the exact version of the relation \cref{eq:pci} \footnote{Signs of cuts are purely a matter of convention related to orientation, and the alternating signs of the contour relation merely indicate a different convention. Our picture of hyperplanes gives a natural relation of orientations.}, before simplifying $\pi \csc(\pi \epsilon)$ to $1/\epsilon$ and dropping the higher-order cuts by working modulo $i\pi$.

Thus a simple picture of one-loop cut relations emerges from the combinatorics of regions in parameter space. Beyond one loop, $F$ can exhibit more complicated behavior involving factorization or self-intersection, and as mentioned in the previous section, cuts are not always unique. Analytic continuation may change the number of  integration regions. Nevertheless, there are indications of similarly visible cut relations in simple multiloop cases. These issues and results will be presented in forthcoming work.

\section*{Summary and discussion}

In parametric space, cut integrals may be defined using integration domains bounded by the second Symanzik polynomial and the coordinate hyperplanes complementary to the set of cut edges.  Maximal cuts are integrated over domains bounded only by the second Symanzik polynomial. 
These integral representations may further expand the utility of parametric techniques, including numerical treatments, and may simplify some aspects of analytic continuation by focusing on the graph polynomials. 
It would be interesting to consider these integration domains in the Lee-Pomeransky representation \cite{Lee:2013hzt}
and the Baikov representation \cite{Cutkosky:1960sp,Baikov:1996iu}.

A complete treatment of discontinuities should include second-type singularities involving the infinite limit of loop momentum. For one-loop integrals, the decomposition theorem relates them to first-type singularities described by cutting edges, and for simpler multiloop integrals there are apparently similar relations. Second-type singularities are associated to $\cU=0$ \cite{10.1063/1.1724262,NakanishiGraphTheoryFeynman}, so  this condition may play some role as a boundary as well. Alternatively, compactifying the full integration domain might improve the analysis of regions.

By using cut contours defined by their boundaries, we have been able to give a simple inclusion-exclusion picture underlying the relation of \cref{eq:pci}, lifting the relation to an equality rather than a congruence.  It would be interesting to investigate similar relations at higher loop order, to see to what extent the one-loop relations can be applied to subgraphs, and to understand cut relations for non-polylogarithmic Feynman integrals such as the ones found in \cite{Hidding:2021xot}. 
More generally, the graphical structure of the Symanzik polynomials \cite{Bogner:2010kv} could lead to recursive relations among cut integrals.

Our construction was inspired by the period matrix of generalized hypergeometric functions, for which a basis of integration contours 
that are dual to a basis of integrands related by integration-by-parts identities 
leads to an algebraic coaction \cite{Brown:2019jng,Abreu:2019wzk}. It will be interesting to see if this representation of cut integrals sheds light on the diagrammatic coaction for Feynman integrals \cite{Abreu:2017enx,Abreu:2017mtm,Abreu:2021vhb},
 which is based on a similar duality, or on the representation of Feynman integrals as generalized hypergeometric functions.

More broadly, the idea of identifying discontinuities as integrals over full-dimensional algebraic varieties might find application in other classes of functions, such as complete amplitudes in which some of the singularities of individual Feynman integrals have canceled out. 
In contexts where loop amplitudes are computed as volumes of some geometry (e.g. \cite{Schnetz:2010pd,Davydychev:1997wa, Arkani-Hamed:2014dca}), it would be interesting to see whether cuts are also volumes, and to explore the nature of those geometries.

\begin{acknowledgments}
It is a pleasure to thank Samuel Abreu, Nima Arkani-Hamed, Michael Borinsky, Claude Duhr, Einan Gardi, Riccardo Gonzo, Hofie Hannesdottir, Song He, Martijn Hidding, Sebastian Mizera, and Matteo Parisi for stimulating discussions and enlightening suggestions.

This research was supported by the Munich Institute for Astro-, Particle and BioPhysics (MIAPbP) which is funded by the Deutsche Forschungsgemeinschaft (DFG, German Research Foundation) under Germany's Excellence Strategy---EXC-2094---390783311, and by the J.~Robert Oppenheimer Visiting Professorship at the Institute for Advanced Study. %
\end{acknowledgments}

\bibliography{referencesFP}

\begin{thebibliography}{34}%
\makeatletter
\providecommand \@ifxundefined [1]{%
 \@ifx{#1\undefined}
}%
\providecommand \@ifnum [1]{%
 \ifnum #1\expandafter \@firstoftwo
 \else \expandafter \@secondoftwo
 \fi
}%
\providecommand \@ifx [1]{%
 \ifx #1\expandafter \@firstoftwo
 \else \expandafter \@secondoftwo
 \fi
}%
\providecommand \natexlab [1]{#1}%
\providecommand \enquote  [1]{``#1''}%
\providecommand \bibnamefont  [1]{#1}%
\providecommand \bibfnamefont [1]{#1}%
\providecommand \citenamefont [1]{#1}%
\providecommand \href@noop [0]{\@secondoftwo}%
\providecommand \href [0]{\begingroup \@sanitize@url \@href}%
\providecommand \@href[1]{\@@startlink{#1}\@@href}%
\providecommand \@@href[1]{\endgroup#1\@@endlink}%
\providecommand \@sanitize@url [0]{\catcode `\\12\catcode `\$12\catcode
  `\&12\catcode `\#12\catcode `\^12\catcode `\_12\catcode `\%12\relax}%
\providecommand \@@startlink[1]{}%
\providecommand \@@endlink[0]{}%
\providecommand \url  [0]{\begingroup\@sanitize@url \@url }%
\providecommand \@url [1]{\endgroup\@href {#1}{\urlprefix }}%
\providecommand \urlprefix  [0]{URL }%
\providecommand \Eprint [0]{\href }%
\providecommand \doibase [0]{http://dx.doi.org/}%
\providecommand \selectlanguage [0]{\@gobble}%
\providecommand \bibinfo  [0]{\@secondoftwo}%
\providecommand \bibfield  [0]{\@secondoftwo}%
\providecommand \translation [1]{[#1]}%
\providecommand \BibitemOpen [0]{}%
\providecommand \bibitemStop [0]{}%
\providecommand \bibitemNoStop [0]{.\EOS\space}%
\providecommand \EOS [0]{\spacefactor3000\relax}%
\providecommand \BibitemShut  [1]{\csname bibitem#1\endcsname}%
\let\auto@bib@innerbib\@empty
\bibitem [{Note1()}]{Note1}%
  \BibitemOpen
  \bibinfo {note} {We suppress the factor $e^{L \gamma _E \epsilon }$ that is
  conventional in dimensional regularization.}\BibitemShut {Stop}%
\bibitem [{\citenamefont {Landau}(1959)}]{Landau:1959fi}%
  \BibitemOpen
  \bibfield  {author} {\bibinfo {author} {\bibfnamefont {L.~D.}\ \bibnamefont
  {Landau}},\ }\href {\doibase 10.1016/B978-0-08-010586-4.50103-6} {\bibfield
  {journal} {\bibinfo  {journal} {Nucl. Phys.}\ }\textbf {\bibinfo {volume}
  {13}},\ \bibinfo {pages} {181} (\bibinfo {year} {1959})}\BibitemShut
  {NoStop}%
\bibitem [{\citenamefont {Eden}\ \emph {et~al.}(1966)\citenamefont {Eden},
  \citenamefont {Landshoff}, \citenamefont {Olive},\ and\ \citenamefont
  {Polkinghorne}}]{SMatrix}%
  \BibitemOpen
  \bibfield  {author} {\bibinfo {author} {\bibfnamefont {R.}~\bibnamefont
  {Eden}}, \bibinfo {author} {\bibfnamefont {P.}~\bibnamefont {Landshoff}},
  \bibinfo {author} {\bibfnamefont {D.}~\bibnamefont {Olive}}, \ and\ \bibinfo
  {author} {\bibfnamefont {J.}~\bibnamefont {Polkinghorne}},\ }\href@noop {}
  {\emph {\bibinfo {title} {The Analytic S-Matrix}}}\ (\bibinfo  {publisher}
  {Cambridge at the University Press},\ \bibinfo {year} {1966})\BibitemShut
  {NoStop}%
\bibitem [{\citenamefont {Cutkosky}(1960)}]{Cutkosky:1960sp}%
  \BibitemOpen
  \bibfield  {author} {\bibinfo {author} {\bibfnamefont {R.~E.}\ \bibnamefont
  {Cutkosky}},\ }\href {\doibase 10.1063/1.1703676} {\bibfield  {journal}
  {\bibinfo  {journal} {J. Math. Phys.}\ }\textbf {\bibinfo {volume} {1}},\
  \bibinfo {pages} {429} (\bibinfo {year} {1960})}\BibitemShut {NoStop}%
\bibitem [{\citenamefont {Hwa}\ and\ \citenamefont {Teplitz}(1966)}]{teplitz}%
  \BibitemOpen
  \bibfield  {author} {\bibinfo {author} {\bibfnamefont {R.~C.}\ \bibnamefont
  {Hwa}}\ and\ \bibinfo {author} {\bibfnamefont {V.~L.}\ \bibnamefont
  {Teplitz}},\ }\href@noop {} {\emph {\bibinfo {title} {{Homology and Feynman
  integrals}}}}\ (\bibinfo  {publisher} {W. A. Benjamin Inc.},\ \bibinfo {year}
  {1966})\BibitemShut {NoStop}%
\bibitem [{Note2()}]{Note2}%
  \BibitemOpen
  \bibinfo {note} {For subleading singularities with $J \protect \neq E$ and
  $\protect \mathcal {U}=0$, the second equation is satisfied trivially but can
  be refined to expose the singularity. See for example \cite
  {SMatrix,NakanishiGraphTheoryFeynman,Boyling1968AHA,Klausen:2021yrt,Hannesdottir:2022bmo}}\BibitemShut
  {NoStop}%
\bibitem [{\citenamefont {Abreu}\ \emph
  {et~al.}(2017{\natexlab{a}})\citenamefont {Abreu}, \citenamefont {Britto},
  \citenamefont {Duhr},\ and\ \citenamefont {Gardi}}]{Abreu:2017ptx}%
  \BibitemOpen
  \bibfield  {author} {\bibinfo {author} {\bibfnamefont {S.}~\bibnamefont
  {Abreu}}, \bibinfo {author} {\bibfnamefont {R.}~\bibnamefont {Britto}},
  \bibinfo {author} {\bibfnamefont {C.}~\bibnamefont {Duhr}}, \ and\ \bibinfo
  {author} {\bibfnamefont {E.}~\bibnamefont {Gardi}},\ }\href {\doibase
  10.1007/JHEP06(2017)114} {\bibfield  {journal} {\bibinfo  {journal} {JHEP}\
  }\textbf {\bibinfo {volume} {06}},\ \bibinfo {pages} {114} (\bibinfo {year}
  {2017}{\natexlab{a}})},\ \Eprint {http://arxiv.org/abs/1702.03163}
  {arXiv:1702.03163 [hep-th]} \BibitemShut {NoStop}%
\bibitem [{\citenamefont {Aomoto}\ and\ \citenamefont
  {Kita}(2011)}]{AomotoKita}%
  \BibitemOpen
  \bibfield  {author} {\bibinfo {author} {\bibfnamefont {K.}~\bibnamefont
  {Aomoto}}\ and\ \bibinfo {author} {\bibfnamefont {M.}~\bibnamefont {Kita}},\
  }\href@noop {} {\emph {\bibinfo {title} {Theory of Hypergeometric
  Functions}}},\ Springer Monographs in Mathematics\ (\bibinfo  {publisher}
  {Springer Japan},\ \bibinfo {year} {2011})\BibitemShut {NoStop}%
\bibitem [{\citenamefont {Fotiadi}\ \emph {et~al.}(1965)\citenamefont
  {Fotiadi}, \citenamefont {Froissart}, \citenamefont {Lascoux},\ and\
  \citenamefont {Pham}}]{fotiadi}%
  \BibitemOpen
  \bibfield  {author} {\bibinfo {author} {\bibfnamefont {D.}~\bibnamefont
  {Fotiadi}}, \bibinfo {author} {\bibfnamefont {M.}~\bibnamefont {Froissart}},
  \bibinfo {author} {\bibfnamefont {J.}~\bibnamefont {Lascoux}}, \ and\
  \bibinfo {author} {\bibfnamefont {F.}~\bibnamefont {Pham}},\ }\href@noop {}
  {\bibfield  {journal} {\bibinfo  {journal} {Topology}\ }\textbf {\bibinfo
  {volume} {4}},\ \bibinfo {pages} {159} (\bibinfo {year} {1965})}\BibitemShut
  {NoStop}%
\bibitem [{\citenamefont {Pham}(2005)}]{PhamBook}%
  \BibitemOpen
  \bibinfo {editor} {\bibfnamefont {F.}~\bibnamefont {Pham}},\ ed.,\ \href@noop
  {} {\emph {\bibinfo {title} {{Singularities of Integrals}}}}\ (\bibinfo
  {publisher} {Springer},\ \bibinfo {year} {2005})\BibitemShut {NoStop}%
\bibitem [{\citenamefont {Hannesdottir}\ \emph {et~al.}(2022)\citenamefont
  {Hannesdottir}, \citenamefont {McLeod}, \citenamefont {Schwartz},\ and\
  \citenamefont {Vergu}}]{Hannesdottir:2022xki}%
  \BibitemOpen
  \bibfield  {author} {\bibinfo {author} {\bibfnamefont {H.~S.}\ \bibnamefont
  {Hannesdottir}}, \bibinfo {author} {\bibfnamefont {A.~J.}\ \bibnamefont
  {McLeod}}, \bibinfo {author} {\bibfnamefont {M.~D.}\ \bibnamefont
  {Schwartz}}, \ and\ \bibinfo {author} {\bibfnamefont {C.}~\bibnamefont
  {Vergu}},\ }\href@noop {} {\  (\bibinfo {year} {2022})},\ \Eprint
  {http://arxiv.org/abs/2211.07633} {arXiv:2211.07633 [hep-th]} \BibitemShut
  {NoStop}%
\bibitem [{\citenamefont {Berghoff}\ and\ \citenamefont
  {Panzer}(2022)}]{Berghoff:2022mqu}%
  \BibitemOpen
  \bibfield  {author} {\bibinfo {author} {\bibfnamefont {M.}~\bibnamefont
  {Berghoff}}\ and\ \bibinfo {author} {\bibfnamefont {E.}~\bibnamefont
  {Panzer}},\ }\href@noop {} {\  (\bibinfo {year} {2022})},\ \Eprint
  {http://arxiv.org/abs/2212.06661} {arXiv:2212.06661 [math-ph]} \BibitemShut
  {NoStop}%
\bibitem [{Note3()}]{Note3}%
  \BibitemOpen
  \bibinfo {note} {At the anomalous threshold, $F$ is tangent to $H_3$ away
  from $\Delta $.}\BibitemShut {Stop}%
\bibitem [{Note4()}]{Note4}%
  \BibitemOpen
  \bibinfo {note} {This choice ensures that cuts agree with physical
  discontinuities. A different choice would give a linear combination of cuts
  with additional on-shell conditions. For the maximal cut, the validity of our
  choice can be seen by consistency with the case of $n=2$, where the maximal
  cut is also a momentum discontinuity.}\BibitemShut {Stop}%
\bibitem [{Note5()}]{Note5}%
  \BibitemOpen
  \bibinfo {note} {This formula differs from the conventions of \cite
  {Abreu:2017ptx} in that we have not removed overall factors of $2\pi
  i$.}\BibitemShut {Stop}%
\bibitem [{\citenamefont {Abreu}\ \emph {et~al.}(2021)\citenamefont {Abreu},
  \citenamefont {Britto}, \citenamefont {Duhr}, \citenamefont {Gardi},\ and\
  \citenamefont {Matthew}}]{Abreu:2021vhb}%
  \BibitemOpen
  \bibfield  {author} {\bibinfo {author} {\bibfnamefont {S.}~\bibnamefont
  {Abreu}}, \bibinfo {author} {\bibfnamefont {R.}~\bibnamefont {Britto}},
  \bibinfo {author} {\bibfnamefont {C.}~\bibnamefont {Duhr}}, \bibinfo {author}
  {\bibfnamefont {E.}~\bibnamefont {Gardi}}, \ and\ \bibinfo {author}
  {\bibfnamefont {J.}~\bibnamefont {Matthew}},\ }\href {\doibase
  10.1007/JHEP10(2021)131} {\bibfield  {journal} {\bibinfo  {journal} {JHEP}\
  }\textbf {\bibinfo {volume} {10}},\ \bibinfo {pages} {131} (\bibinfo {year}
  {2021})},\ \Eprint {http://arxiv.org/abs/2106.01280} {arXiv:2106.01280
  [hep-th]} \BibitemShut {NoStop}%
\bibitem [{Note6()}]{Note6}%
  \BibitemOpen
  \bibinfo {note} {These additional powers of $i\pi $ are conventionally
  removed through normalization when studying the functional form of
  cuts.}\BibitemShut {Stop}%
\bibitem [{Note7()}]{Note7}%
  \BibitemOpen
  \bibinfo {note} {Signs of cuts are purely a matter of convention related to
  orientation, and the alternating signs of the contour relation merely
  indicate a different convention. Our picture of hyperplanes gives a natural
  relation of orientations.}\BibitemShut {Stop}%
\bibitem [{\citenamefont {Lee}\ and\ \citenamefont
  {Pomeransky}(2013)}]{Lee:2013hzt}%
  \BibitemOpen
  \bibfield  {author} {\bibinfo {author} {\bibfnamefont {R.~N.}\ \bibnamefont
  {Lee}}\ and\ \bibinfo {author} {\bibfnamefont {A.~A.}\ \bibnamefont
  {Pomeransky}},\ }\href {\doibase 10.1007/JHEP11(2013)165} {\bibfield
  {journal} {\bibinfo  {journal} {JHEP}\ }\textbf {\bibinfo {volume} {11}},\
  \bibinfo {pages} {165} (\bibinfo {year} {2013})},\ \Eprint
  {http://arxiv.org/abs/1308.6676} {arXiv:1308.6676 [hep-ph]} \BibitemShut
  {NoStop}%
\bibitem [{\citenamefont {Baikov}(1997)}]{Baikov:1996iu}%
  \BibitemOpen
  \bibfield  {author} {\bibinfo {author} {\bibfnamefont {P.~A.}\ \bibnamefont
  {Baikov}},\ }\href {\doibase 10.1016/S0168-9002(97)00126-5} {\bibfield
  {journal} {\bibinfo  {journal} {Nucl. Instrum. Meth. A}\ }\textbf {\bibinfo
  {volume} {389}},\ \bibinfo {pages} {347} (\bibinfo {year} {1997})},\ \Eprint
  {http://arxiv.org/abs/hep-ph/9611449} {arXiv:hep-ph/9611449} \BibitemShut
  {NoStop}%
\bibitem [{\citenamefont {Fairlie}\ \emph {et~al.}(2004)\citenamefont
  {Fairlie}, \citenamefont {Landshoff}, \citenamefont {Nuttall},\ and\
  \citenamefont {Polkinghorne}}]{10.1063/1.1724262}%
  \BibitemOpen
  \bibfield  {author} {\bibinfo {author} {\bibfnamefont {D.~B.}\ \bibnamefont
  {Fairlie}}, \bibinfo {author} {\bibfnamefont {P.~V.}\ \bibnamefont
  {Landshoff}}, \bibinfo {author} {\bibfnamefont {J.}~\bibnamefont {Nuttall}},
  \ and\ \bibinfo {author} {\bibfnamefont {J.~C.}\ \bibnamefont
  {Polkinghorne}},\ }\href {\doibase 10.1063/1.1724262} {\bibfield  {journal}
  {\bibinfo  {journal} {Journal of Mathematical Physics}\ }\textbf {\bibinfo
  {volume} {3}},\ \bibinfo {pages} {594} (\bibinfo {year} {2004})},\ \Eprint
  {http://arxiv.org/abs/https://pubs.aip.org/aip/jmp/article-pdf/3/4/594/11042358/594\_1\_online.pdf}
  {https://pubs.aip.org/aip/jmp/article-pdf/3/4/594/11042358/594\_1\_online.pdf}
  \BibitemShut {NoStop}%
\bibitem [{\citenamefont {Nakanishi}(1971)}]{NakanishiGraphTheoryFeynman}%
  \BibitemOpen
  \bibfield  {author} {\bibinfo {author} {\bibfnamefont {N.}~\bibnamefont
  {Nakanishi}},\ }\href@noop {} {\emph {\bibinfo {title} {Graph {{Theory}} and
  {{Feynman Integrals}}}}}\ (\bibinfo  {publisher} {Gordon and Breach},\
  \bibinfo {year} {1971})\BibitemShut {NoStop}%
\bibitem [{\citenamefont {Hidding}(2021)}]{Hidding:2021xot}%
  \BibitemOpen
  \bibfield  {author} {\bibinfo {author} {\bibfnamefont {M.}~\bibnamefont
  {Hidding}},\ }\emph {\bibinfo {title} {{Computational and mathematical
  aspects of Feynman integrals}}},\ \href@noop {} {Ph.D. thesis},\ \bibinfo
  {school} {Trinity Coll., Dublin} (\bibinfo {year} {2021})\BibitemShut
  {NoStop}%
\bibitem [{\citenamefont {Bogner}\ and\ \citenamefont
  {Weinzierl}(2010)}]{Bogner:2010kv}%
  \BibitemOpen
  \bibfield  {author} {\bibinfo {author} {\bibfnamefont {C.}~\bibnamefont
  {Bogner}}\ and\ \bibinfo {author} {\bibfnamefont {S.}~\bibnamefont
  {Weinzierl}},\ }\href {\doibase 10.1142/S0217751X10049438} {\bibfield
  {journal} {\bibinfo  {journal} {Int. J. Mod. Phys. A}\ }\textbf {\bibinfo
  {volume} {25}},\ \bibinfo {pages} {2585} (\bibinfo {year} {2010})},\ \Eprint
  {http://arxiv.org/abs/1002.3458} {arXiv:1002.3458 [hep-ph]} \BibitemShut
  {NoStop}%
\bibitem [{\citenamefont {Brown}\ and\ \citenamefont
  {Dupont}(2019)}]{Brown:2019jng}%
  \BibitemOpen
  \bibfield  {author} {\bibinfo {author} {\bibfnamefont {F.}~\bibnamefont
  {Brown}}\ and\ \bibinfo {author} {\bibfnamefont {C.}~\bibnamefont {Dupont}},\
  }\href@noop {} {\  (\bibinfo {year} {2019})},\ \Eprint
  {http://arxiv.org/abs/1907.06603} {arXiv:1907.06603 [math.AG]} \BibitemShut
  {NoStop}%
\bibitem [{\citenamefont {Abreu}\ \emph {et~al.}(2020)\citenamefont {Abreu},
  \citenamefont {Britto}, \citenamefont {Duhr}, \citenamefont {Gardi},\ and\
  \citenamefont {Matthew}}]{Abreu:2019wzk}%
  \BibitemOpen
  \bibfield  {author} {\bibinfo {author} {\bibfnamefont {S.}~\bibnamefont
  {Abreu}}, \bibinfo {author} {\bibfnamefont {R.}~\bibnamefont {Britto}},
  \bibinfo {author} {\bibfnamefont {C.}~\bibnamefont {Duhr}}, \bibinfo {author}
  {\bibfnamefont {E.}~\bibnamefont {Gardi}}, \ and\ \bibinfo {author}
  {\bibfnamefont {J.}~\bibnamefont {Matthew}},\ }\href {\doibase
  10.1007/JHEP02(2020)122} {\bibfield  {journal} {\bibinfo  {journal} {JHEP}\
  }\textbf {\bibinfo {volume} {02}},\ \bibinfo {pages} {122} (\bibinfo {year}
  {2020})},\ \Eprint {http://arxiv.org/abs/1910.08358} {arXiv:1910.08358
  [hep-th]} \BibitemShut {NoStop}%
\bibitem [{\citenamefont {Abreu}\ \emph
  {et~al.}(2017{\natexlab{b}})\citenamefont {Abreu}, \citenamefont {Britto},
  \citenamefont {Duhr},\ and\ \citenamefont {Gardi}}]{Abreu:2017enx}%
  \BibitemOpen
  \bibfield  {author} {\bibinfo {author} {\bibfnamefont {S.}~\bibnamefont
  {Abreu}}, \bibinfo {author} {\bibfnamefont {R.}~\bibnamefont {Britto}},
  \bibinfo {author} {\bibfnamefont {C.}~\bibnamefont {Duhr}}, \ and\ \bibinfo
  {author} {\bibfnamefont {E.}~\bibnamefont {Gardi}},\ }\href {\doibase
  10.1103/PhysRevLett.119.051601} {\bibfield  {journal} {\bibinfo  {journal}
  {Phys. Rev. Lett.}\ }\textbf {\bibinfo {volume} {119}},\ \bibinfo {pages}
  {051601} (\bibinfo {year} {2017}{\natexlab{b}})},\ \Eprint
  {http://arxiv.org/abs/1703.05064} {arXiv:1703.05064 [hep-th]} \BibitemShut
  {NoStop}%
\bibitem [{\citenamefont {Abreu}\ \emph
  {et~al.}(2017{\natexlab{c}})\citenamefont {Abreu}, \citenamefont {Britto},
  \citenamefont {Duhr},\ and\ \citenamefont {Gardi}}]{Abreu:2017mtm}%
  \BibitemOpen
  \bibfield  {author} {\bibinfo {author} {\bibfnamefont {S.}~\bibnamefont
  {Abreu}}, \bibinfo {author} {\bibfnamefont {R.}~\bibnamefont {Britto}},
  \bibinfo {author} {\bibfnamefont {C.}~\bibnamefont {Duhr}}, \ and\ \bibinfo
  {author} {\bibfnamefont {E.}~\bibnamefont {Gardi}},\ }\href {\doibase
  10.1007/JHEP12(2017)090} {\bibfield  {journal} {\bibinfo  {journal} {JHEP}\
  }\textbf {\bibinfo {volume} {12}},\ \bibinfo {pages} {090} (\bibinfo {year}
  {2017}{\natexlab{c}})},\ \Eprint {http://arxiv.org/abs/1704.07931}
  {arXiv:1704.07931 [hep-th]} \BibitemShut {NoStop}%
\bibitem [{\citenamefont {Schnetz}(2010)}]{Schnetz:2010pd}%
  \BibitemOpen
  \bibfield  {author} {\bibinfo {author} {\bibfnamefont {O.}~\bibnamefont
  {Schnetz}},\ }\href@noop {} {\  (\bibinfo {year} {2010})},\ \Eprint
  {http://arxiv.org/abs/1010.5334} {arXiv:1010.5334 [hep-th]} \BibitemShut
  {NoStop}%
\bibitem [{\citenamefont {Davydychev}\ and\ \citenamefont
  {Delbourgo}(1998)}]{Davydychev:1997wa}%
  \BibitemOpen
  \bibfield  {author} {\bibinfo {author} {\bibfnamefont {A.~I.}\ \bibnamefont
  {Davydychev}}\ and\ \bibinfo {author} {\bibfnamefont {R.}~\bibnamefont
  {Delbourgo}},\ }\href {\doibase 10.1063/1.532513} {\bibfield  {journal}
  {\bibinfo  {journal} {J. Math. Phys.}\ }\textbf {\bibinfo {volume} {39}},\
  \bibinfo {pages} {4299} (\bibinfo {year} {1998})},\ \Eprint
  {http://arxiv.org/abs/hep-th/9709216} {arXiv:hep-th/9709216} \BibitemShut
  {NoStop}%
\bibitem [{\citenamefont {Arkani-Hamed}\ \emph {et~al.}(2015)\citenamefont
  {Arkani-Hamed}, \citenamefont {Hodges},\ and\ \citenamefont
  {Trnka}}]{Arkani-Hamed:2014dca}%
  \BibitemOpen
  \bibfield  {author} {\bibinfo {author} {\bibfnamefont {N.}~\bibnamefont
  {Arkani-Hamed}}, \bibinfo {author} {\bibfnamefont {A.}~\bibnamefont
  {Hodges}}, \ and\ \bibinfo {author} {\bibfnamefont {J.}~\bibnamefont
  {Trnka}},\ }\href {\doibase 10.1007/JHEP08(2015)030} {\bibfield  {journal}
  {\bibinfo  {journal} {JHEP}\ }\textbf {\bibinfo {volume} {08}},\ \bibinfo
  {pages} {030} (\bibinfo {year} {2015})},\ \Eprint
  {http://arxiv.org/abs/1412.8478} {arXiv:1412.8478 [hep-th]} \BibitemShut
  {NoStop}%
\bibitem [{\citenamefont {Boyling}(1968)}]{Boyling1968AHA}%
  \BibitemOpen
  \bibfield  {author} {\bibinfo {author} {\bibfnamefont {J.~B.}\ \bibnamefont
  {Boyling}},\ }\href@noop {} {\bibfield  {journal} {\bibinfo  {journal} {Il
  Nuovo Cimento A (1965-1970)}\ }\textbf {\bibinfo {volume} {53}},\ \bibinfo
  {pages} {351} (\bibinfo {year} {1968})}\BibitemShut {NoStop}%
\bibitem [{\citenamefont {Klausen}(2022)}]{Klausen:2021yrt}%
  \BibitemOpen
  \bibfield  {author} {\bibinfo {author} {\bibfnamefont {R.~P.}\ \bibnamefont
  {Klausen}},\ }\href {\doibase 10.1007/JHEP02(2022)004} {\bibfield  {journal}
  {\bibinfo  {journal} {JHEP}\ }\textbf {\bibinfo {volume} {02}},\ \bibinfo
  {pages} {004} (\bibinfo {year} {2022})},\ \Eprint
  {http://arxiv.org/abs/2109.07584} {arXiv:2109.07584 [hep-th]} \BibitemShut
  {NoStop}%
\bibitem [{\citenamefont {Hannesdottir}\ and\ \citenamefont
  {Mizera}(2023)}]{Hannesdottir:2022bmo}%
  \BibitemOpen
  \bibfield  {author} {\bibinfo {author} {\bibfnamefont {H.~S.}\ \bibnamefont
  {Hannesdottir}}\ and\ \bibinfo {author} {\bibfnamefont {S.}~\bibnamefont
  {Mizera}},\ }\href {\doibase 10.1007/978-3-031-18258-7} {\emph {\bibinfo
  {title} {{What is the i\ensuremath{\varepsilon} for the S-matrix?}}}},\
  SpringerBriefs in Physics\ (\bibinfo  {publisher} {Springer},\ \bibinfo
  {year} {2023})\ \Eprint {http://arxiv.org/abs/2204.02988} {arXiv:2204.02988
  [hep-th]} \BibitemShut {NoStop}%
\end{thebibliography}%
  
\end{document}